# Measuring Domain Knowledge for Early Prediction of Student Performance: A Semantic Approach

Anupam Khan
Computer Science and Engineering
Indian Institute of Technology
Kharagpur, India
anupamkh@iitkgp.ac.in

Sourav Ghosh
Computer Science and Engineering
Indian Institute of Technology
Kharagpur, India
sourav.ghosh@outlook.in

Soumya K. Ghosh
Computer Science and Engineering
Indian Institute of Technology
Kharagpur, India
skg@cse.iitkgp.ac.in

*Abstract*—The growing popularity of data mining catalyses the researchers to explore various exciting aspects of education. Early prediction of student performance is an emerging area among them. The researchers have used various predictors in performance modelling studies. Although prior cognition can affect student performance, establishing their relationship is still an open research challenge. Quantifying the knowledge from readily available data is the major challenge here. We have proposed a semantic approach for this purpose. Association mining on nearly 0.35 million observations establishes that prior cognition impacts the student performance. The proposed approach of measuring domain knowledge can help the early performance modelling studies to use it as a predictor.

## I. INTRODUCTION

Student performance modelling is the most popular and challenging research direction in educational data mining (EDM) literature [1], [2]. The researchers have started emphasising on early prediction before course commencement in recent years [3], [4], [5]. The availability of more options and time for guiding the at-risk student is one of the driving force behind that. Nevertheless, the factors, influencing student performance, should be identified first before building up an early prediction model [6], [7]. The performance prediction studies have primarily used past performance, internal and external assessment as predictors [8], [9]. The internal assessment marks turn out to be most effective among them [10]. However, these marks are available only during the tenure of the course. It is, therefore, essential to identify some other predictors that are available before course commencement. The cognitive ability can influence student performance [11]. The knowledge on the course topic, therefore, is an exciting factor which can be analysed further [12].

According to Bloom's taxonomy [13], knowledge involves the recall or recognition of terms, ideas, processes, and theories. The knowledge in a similar topic can be referred to as the domain knowledge in the learning context. However, quantifying the domain knowledge is a research challenge [14]. Importantly, the EDM researchers have widely explored the regression and classification methods on structured data for student performance analysis [9], [5]. Significantly, real-life data are not structured always. The unstructured data, like course syllabus, could be useful to extract valuable information for analysing various aspects of the learning process [15]. A few studies [16], [17], [18] measure the domain knowledge based on previous experience, specialisation of study, questionnaires and pre-test score. However, the proposed methods in these studies may not be suitable for quantifying the knowledge automatically in EDM context. It is essential to mention here that the course syllabus is available in digital format nowadays, which contains valuable information about the course content. The course content, in fact, represents the repositories and cognitive models of the knowledge components to be learned [1]. The primary objectives of the present study, therefore, are as follows:

- To propose a mechanism of measuring domain knowledge from existing academic records.
- Establishing the association between measured domain knowledge and student performance.

Although the researchers have widely used text mining in other domains, [19] have noticed little use of it for understanding the learning processes in education. The authors in [5] have also noticed the similar trend in student performance analysis related studies of recent times. Importantly, the traditional cosine similarity-based approaches of mining unstructured data have its' own limitations. The application of semantics in text mining has created a new avenue to overcome those. For example, [20] has used semantic approaches for identifying duplicate questions in the online discussion forum of e-learning tools. In a study [21], the authors have utilised it for characterising users based on the message similarity in online discussion forum. Besides this, [22] have used it for measuring the similarity of learning materials as well. In another work [23], the authors have used it for analysing student feedback on teaching. Motivated by these studies, the present work first applies semantic method on unstructured syllabus data to measure the similarity between various courses. It uses the measured similarity and performance of all previously attended courses for determining the domain knowledge. The application of association rule mining further helps in establishing the positive impact of domain knowledge on student performance.

The paper is organised as follows: Section II first introduces the dataset. Section III presents the methodology for quanti-fying domain knowledge and analysing its' impact on student performance. Section IV presents the results and discusses its implication. Finally, Section V concludes this work.

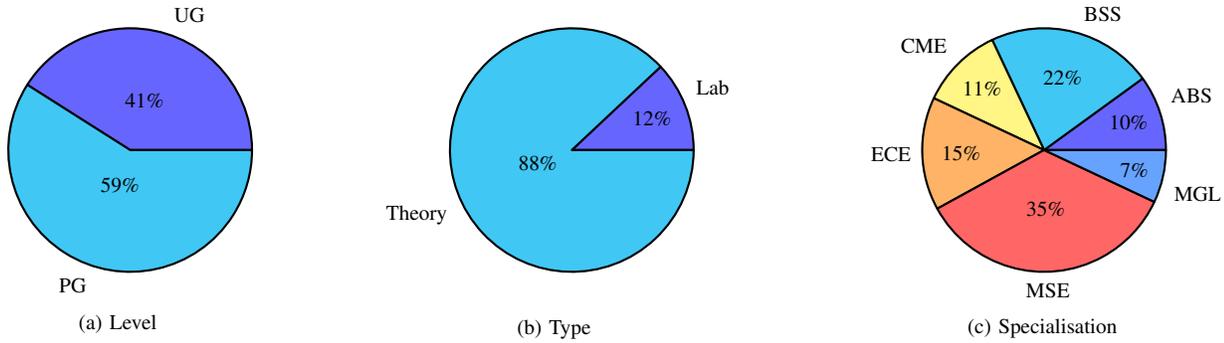

Fig. 1: Variety of courses in first dataset

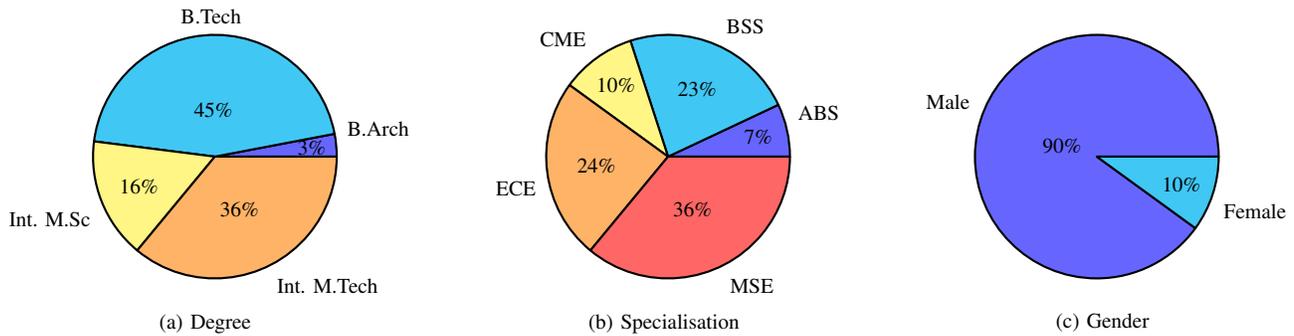

Fig. 2: Variety of students in second dataset

## II. DATASET

This work has used a dataset collected through the academic information management system (AIMS) of an institute of national importance in India (refer as ACAD-INST in this paper). In addition to engineering, ACAD-INST also offers a few basic science courses to the students. The present study has analysed the data of students admitted through a prestigious national level entrance after completing their school level study. It uses two datasets containing relevant information on the course syllabus and student performance for this analysis.

*Course data (first dataset)*: The first dataset contains the information related to all courses offered by various departments of ACAD-INST. A few courses also exist in ACAD-INST for which the syllabus data is not available in AIMS. The present study discards all such courses for the analysis. This dataset contains data of (i) *course-number*, (ii) *course-name*, and (iii) *syllabus* for 2,707 courses.

*Performance Data (second dataset)*: The second dataset maintains the performance record of all students. In ACAD-INST, all students register for the offered courses as per their academic curricula. After completion of the course, the teacher evaluates their performance and upload the grades in AIMS. This dataset contains 492,917 records of 14,965 students. In addition to these, a few course registration records are also available in AIMS for which the syllabus is not available. This dataset discards all those records. It contains the following attributes: (i) *student-code*, (ii) *semester*, (iii) *preceding-CGPA*, (iv) *course-number*, (v) *registration-type*, and (vi) *grade*. The *preceding-CGPA* indicate the CGPA of a student until the preceding semester. For example, the CGPA at the end of the third semester specifies *preceding-CGPA* attribute of all courses registered in the fourth semester. Furthermore, ACAD-INST considers EX, A, B, C, D, P, and F as valid grades. The EX is the highest one, whereas F indicates failure.

Fig. 1 shows the variety of course records in the first dataset. It contains both undergraduate (UG) and postgraduate (PG) level courses, and the dataset is comprised of both theory and laboratory courses as well. These courses belong to six broad specialisations: (i) Agricultural and Biological Science (ABS), (ii) Basic and Social Science (BSS), (iii) Chemical and Materials Engineering (CME), (iv) Electrical and Computational Engineering (ECE), (v) Mechanical and Structural Engineering (MSE), and (vi) Management and Law (MGL). Besides this, Fig. 2 presents the variety of student whose records are available in the second dataset. The second dataset contains records of both male and female students, though the number of female students is comparatively less. These students are enrolled for programs offering any of the following degrees: (i) Bachelor of Architecture (B.Arch), (ii)

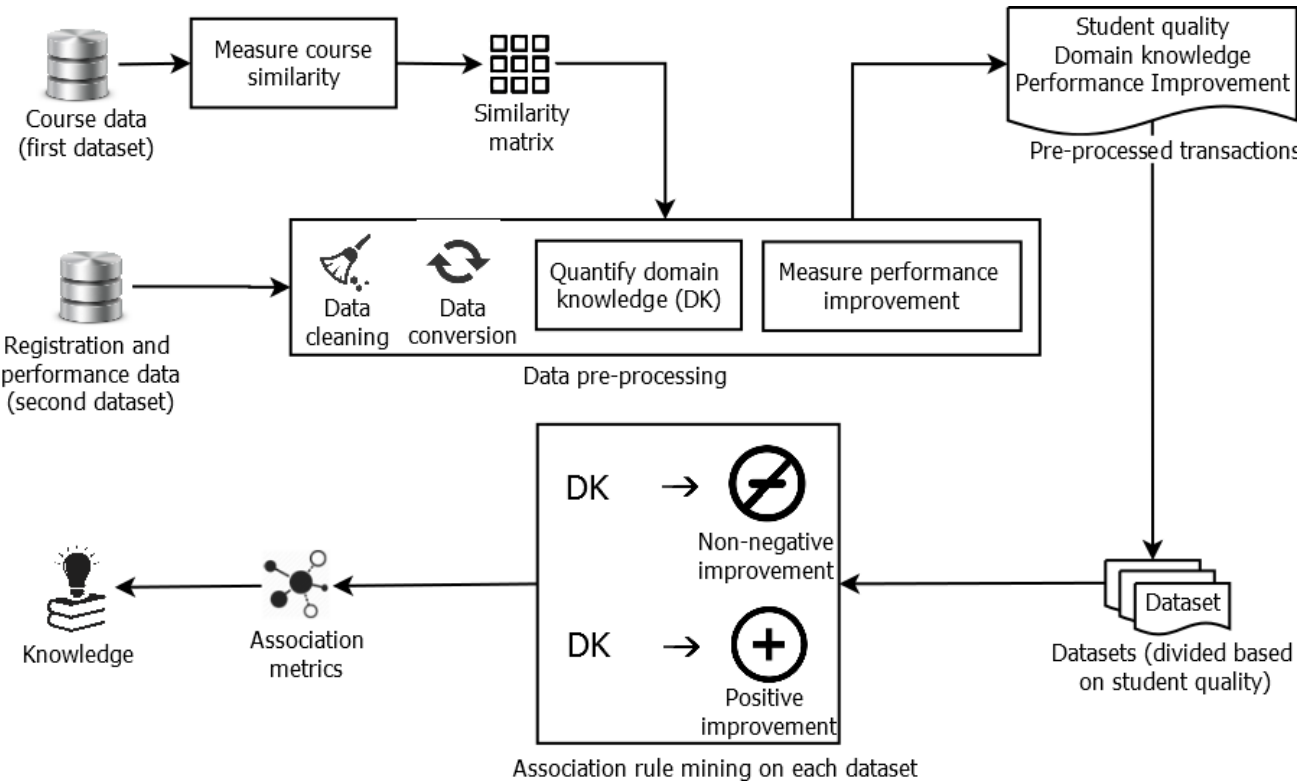

Fig. 3: Overview of student performance analysis

Bachelor of Technology (B.Tech), (iii) Integrated Masters of Science (Int. M.Sc), and (iv) Integrated Masters of Technology (M.Tech). Furthermore, these students belong to any of the five broad specialisations: (i) ABS, (ii) BSS, (iii) CME, (iv) ECE, and (v) MSE.

### III. THE SEMANTIC APPROACH

This section discusses the proposed semantic approach for analysing the influence of domain knowledge on student performance. Fig. 3 graphically presents a brief overview of the proposed approach, where the first and second dataset mentioned in Section II act as the primary input. The application of semantic methods on the first dataset provides the similarity between courses. The data pre-processing step thereafter measures the student quality, domain knowledge and performance improvement for each valid record in the second dataset. The application of association rule mining on the pre-processed dataset finally helps in establishing a significant relationship. The following part of this section elaborates each step in detail.

#### A. Measuring Course Similarity

This subsection discusses the method adapted for computing the semantic similarity between courses and also introduces the course similarity matrix.

*1) Computing semantic similarity:* This study determines the semantic similarity between two courses by comparing their syllabus. Instead of applying traditional cosine similarity approach, the frequent contextual co-occurrence and variation of the words play an essential role here. The constituent phrases of the syllabus in two related subjects do not use the same arrangement of words to indicate a similar or related topic. For example, "graph theory" and "multigraph model" are closely related, but cosine similarity measures them as non-similar. In these cases, simple matching of words using cosine similarity may not be helpful. A metric, which takes the word semantics into account, can help to overcome such issues.

In order to compute the semantic similarity between the syllabus of two courses, this work has used the semantic text similarity model described in [24]. The proposed model calculates the term similarities by using the combination of latent semantic analysis [25] and WordNet [26] knowledge. The semantic similarity measure lies in the range of [0, 1] where 0 indicates non-related courses and 1 indicates exact similarity.

*2) Course similarity matrix:* The semantic similarity measures helps us to build the course similarity matrix here. Significantly, the course syllabus contains some generic terms which do not represent the course topic. These generic terms always produce some similarity between courses. We have, therefore, normalised the similarity measures in order to reduce their effect.

In the next step, it constructs the similarity matrix with the help of similarity measures obtained against each pair of courses. It is a square symmetric matrix of $n_c \times n_c$ dimension where $n_c$ (= 2,707) is the total number of course available in the first dataset. The element ($s_{ij}$) of $i^{th}$ row and $j^{th}$ column is the normalised similarity measure between the $i^{th}$ and $j^{th}$ course. Here, $s_{ij} = s_{ji}$ and all diagonal elements are 1.

### B. Pre-processing the Dataset

In the pre-processing phase, this study first discards all registration records where grade or preceding CGPA is not available. Importantly, the grades are missing in a few records of the second dataset. It is also quite apparent that the preceding CGPA is unavailable in the first semester. The next pre-processing step converts the *grade* attribute of the second dataset to its' numeric equivalent. The letter grades of EX, A, B, C, D, P and F are converted to 10, 9, 8, 7, 6, 5, and 0 respectively. The same conversion formula is used in ACAD-INST as well. The pre-processing phase after that calculates three derived attributes: (i) student quality, (ii) domain knowledge and (iii) performance improvement.

*1) Measuring student quality:* This work classifies the students based on their CGPA until the preceding semester. It uses the *preceding-CGPA* ($c_{t-1}$) attribute of the second dataset to derive the student quality ($S$) here. This classification is carried out based on the following range of preceding CGPA: (i) Extremely Poor (EP): $c_{t-1} < 6$, (ii) Poor (PR): $6 \leq c_{t-1} < 7$, (iii) Medium (MD): $7 \leq c_{t-1} < 8$ (iv) Good (GD): $8 \leq c_{t-1} < 9$, and (v) Very Good (VG): $c_{t-1} \geq 9$. The preceding CGPA varies from 0 to 10 in ACAD-INST. Besides this, students with CGPA greater than or equal to 6 are only eligible to obtain final degree. Therefore, all students below 6 CGPA is considered as extremely poor in this study.

*2) Measuring domain knowledge:* Examination paper is a well-defined instrument to assess the cognition of a student on a specific domain [27]. The performance in similar courses, completed recently, can be a possible indication of knowledge a student possess on the related course topic. The term 'recent' is important here, as it indicates the time dimension. The domain knowledge should be high if a student attends similar courses in recent time, whereas the student can forget similar topics if studied long ago. We have considered the time gap to incorporate the knowledge forgetting aspect. The combination of similarity, performance and time dimension helps us in quantifying the domain knowledge ($\theta$) here.

For each record of filtered second dataset, this study processes the similarity matrix to fetch the semantic similarities of all earlier courses with the present one. It multiplies the numeric equivalent of grades obtained and reciprocal of time gap for each earlier courses with the corresponding similarity measures. The number of semesters between earlier and present course determines the time gap here. This approach produces an array of knowledge factors between the present course and each earlier courses. The nearest integer of maximum knowledge factor is considered as domain knowledge index ($\tilde{\theta}$) in this work which further helps in determining $\theta$. Say, two earlier courses exist with very low and very high similarities respectively. The student performed very poorly in the first one and performed well in the second one. The domain knowledge would be very poor in the first one as the student failed to acquire sufficient knowledge. However, the domain knowledge would be high in case of the second one as the student studied well, acquired knowledge and therefore performed well. Considering the second one for domain knowledge measurement seems be more logical to us. Mathematically, Eq. (1) represents the domain knowledge index for the $i^{th}$ course of a student,

$$\tilde{\theta}_i = \lfloor \max_{\forall j \in P_i} (\frac{s_{ij} \times \tilde{g}_j}{t_{ij}}) \rceil \quad (1)$$

where $P_i$ denotes the set of earlier courses and $\tilde{g}_j$ is the numeric equivalent of the grade obtained in course $j$. The $t_{ij}$ is the time gap between the $i^{th}$ and $j^{th}$ course in terms of semesters. Here, $\tilde{\theta}_i \in \{0, 1, 2, 3, 4, 5, 6, 7, 8, 9, 10\}$, as $0 \leq s_{ij} \leq 1$ and $\tilde{g}_j$ is an integer between 0 and 10 and $t_{ij} \geq 0$.

In the next step, this work determines domain knowledge ($\theta$) by classifying the obtained value of $\tilde{\theta}$. The $\tilde{\theta}$ is classified here as: (i) $\tilde{\theta} \in \{0, 1, 2\}$: Negligible (NG), (ii) $\tilde{\theta} \in \{3, 4\}$: Little (LT), (iii) $\tilde{\theta} \in \{5, 6\}$: Reasonable (RS), (iv) $\tilde{\theta} \in \{7, 8\}$: Sufficient (SF), and (v) $\tilde{\theta} \in \{9, 10\}$: Maximum (MX).

*3) Measuring performance improvement:* Although the grade is a well-accepted measure, it may not be suitable for examining the external performance influencing factors. A weaker student is likely to get a lesser grade than a comparatively superior student. However, the value addition can be an effect of some external factors. In order to measure this value addition, this study has used the improvement factor ($\delta$) as proposed in [28]. It measures the value addition based on the difference between actual and expected grade. The nearest integer of preceding CGPA is considered as the expected grade here.

Based on the value of $\delta$, this study uses two parameters for classifying the improvement. They are termed as (i) Impr1 ($I_1$) and (ii) Impr2 ($I_2$). The $I_1$ indicates whether the performance of a student is as per his or her expectation or not. This work has considered $I_1$ as non-negative in case $\delta \geq 0$ and negative when $\delta < 0$. Furthermore, the $I_2$ determines whether the student has performed beyond expectation. The value of $I_2$ is positive when $\delta > 0$ and non-positive in case $\delta \leq 0$.

*4) Preparing final dataset for analysis:* After obtaining the student quality ($S$), domain knowledge ($\theta$), $I_1$ and $I_2$, this work divides the second dataset into five pre-processed datasets based on the value of $S$. The five pre-processed datasets contain following number of transactions for analysis: (i) Extremely Poor: 22,140, (ii) Poor: 89,107, (iii) Medium: 116,750, (iv) Good: 96,966, and (v) Very Good: 27,576. Further analysis is carried out separately for these five pre-processed datasets.

### C. Association Rule Mining

The objective of this study is to find out the possible association between domain knowledge and performance improvement discussed earlier. This study analyses two probable implications: (i) $\theta \rightarrow I_1$, and (ii) $\theta \rightarrow I_2$. Here $\theta \rightarrow I_1$ implication helps in analysing the possibility of those cases

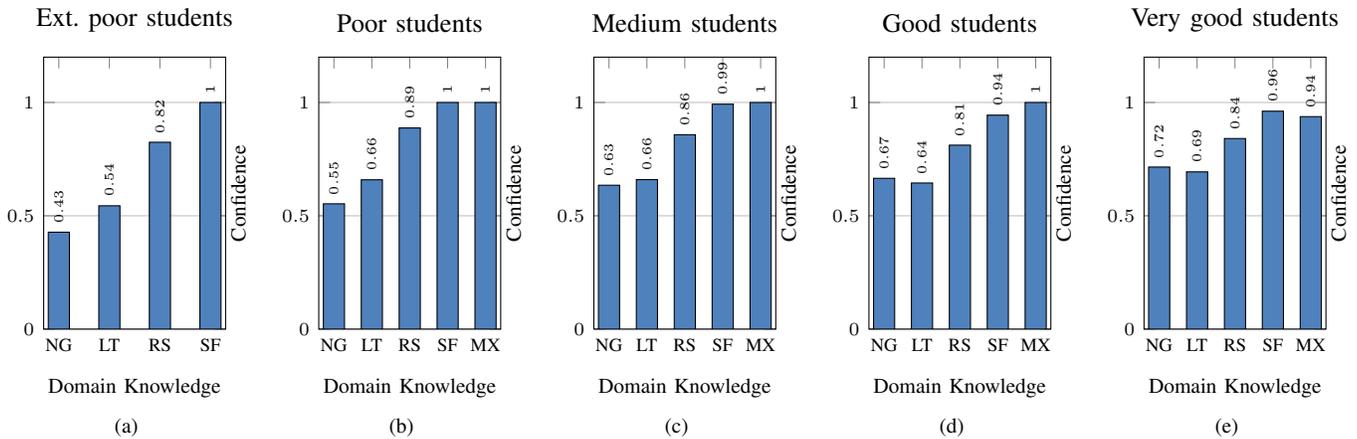

Fig. 4: Confidence of non-negative improvement observed against various domain knowledges (NG: Negligible, LT: Little, RS: Reasonable, SF: Sufficient, and MX: Maximum)

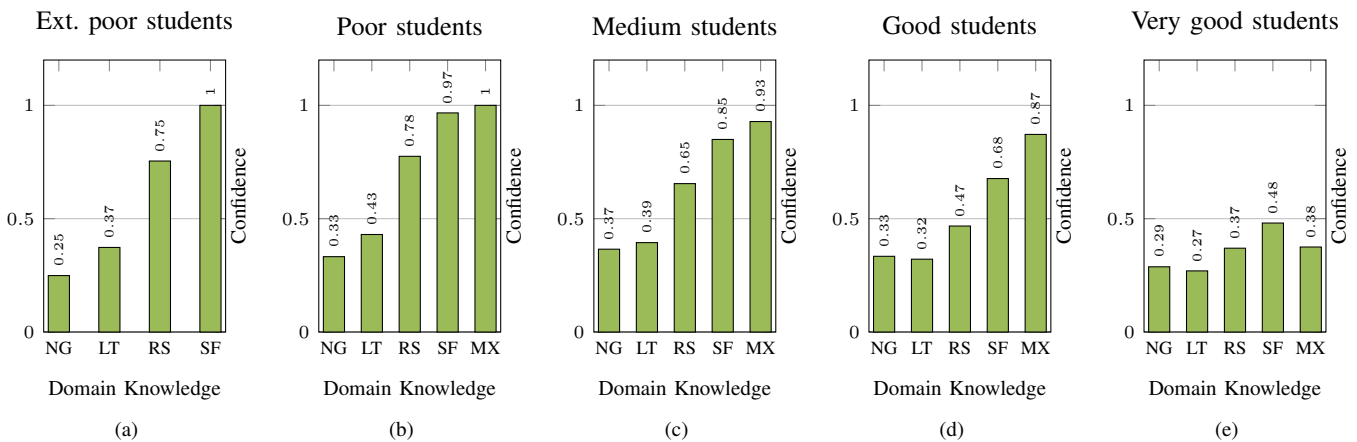

Fig. 5: Confidence of positive improvement observed against various domain knowledges (NG: Negligible, LT: Little, RS: Reasonable, SF: Sufficient, and MX: Maximum)

where specific values of $\theta$ and $I_1$ appear together. Likewise, $\theta \rightarrow I_2$ facilitate a similar analysis between $\theta$ and $I_2$. This work concentrates only on those cases where the student performs either as per expectation or better than that. It, therefore, analyses the association of non-negative and positive improvements in $I_1$ and $I_2$, respectively. It is essential to mention here that the negative improvement in $I_1$ and non-positive improvement in $I_2$ are just complementary to each other. The analysis of the first two should help in understanding the other two complimentary implications as well. In this study, the `arules` package in `R` tool extracts various association metrics from the five datasets separately. This work uses them for establishing the impact of domain knowledge on performance.

## IV. RESULTS AND DISCUSSION

This section presents the results of student performance analysis. Besides this, it also discusses observation and several implications of these results.

### A. Results

In order to understand the impact of domain knowledge on student performance, this study extracts the support and confidence of both $\theta \rightarrow I_1$ and $\theta \rightarrow I_2$ implications. The bar graphs in Fig. 4 and Fig. 5 show all such observed confidences separately for five types of student. Each bar in these two figures presents the confidence of non-negative and positive improvement respectively. Fig. 4a and Fig. 5a contain only four bars as the dataset of extremely poor students does not contain any transaction with maximum domain knowledge. Furthermore, Table I presents the observed support for both non-negative and positive improvements.

The findings show a substantial growth of both positive and non-negative improvement with increasing domain knowledge in general, and the pattern is similar for different qualities of student. It, in turn, implies that the negative improvement is decreasing with growing domain knowledge as well. As an example, Fig. 4a shows that 82% of extremely poor students can perform either as expected or beyond expectation with reasonable domain knowledge. 75% of students are performing beyond expectation in this case (refer to Fig. 5a). However, sufficient domain knowledge enables all extremely poor students to perform beyond expectation. The results in Fig. 4b, 4c, 5b, and 5c indicate that more number of poor and medium quality students can perform either as expected or beyond expectation with growing domain knowledge. The pattern is similar for good students as well (refer to Fig. 4d and Fig. 5d). Fig. 4e and Fig. 5e shows the obtained result for very good students. Although the non-negative and positive improvement increases between little (LT) and sufficient (SF), minor exceptions are observed for negligible (NG) and maximum (MX) domain knowledge. The confidence of positive improvement is also comparatively less in comparison to the other students. In addition to confidence, Table I presents the observed support separately for all qualities of the student. The maximum support of both non-negative and positive improvements is observed for negligible (NG) domain knowledge, and it decreases after that in all cases.

TABLE I. SUPPORT OF NON-NEGATIVE AND POSITIVE IMPROVEMENTS

| Student quality | Domain Knowledge | | | | |
|---|---|---|---|---|---|
| | NG | LT | RS | SF | MX |
| *Improvement: Non-negative* | | | | | |
| Ext. poor | 0.4072 | 0.0247 | 0.0021 | 0.0001 | 0.0000 |
| Poor | 0.4958 | 0.0637 | 0.0051 | 0.0007 | 0.0001 |
| Medium | 0.5069 | 0.1249 | 0.0092 | 0.0011 | 0.0004 |
| Good | 0.4627 | 0.1849 | 0.0127 | 0.0016 | 0.0004 |
| Very good | 0.4311 | 0.2616 | 0.0148 | 0.0018 | 0.0005 |
| *Improvement: Positive* | | | | | |
| Ext. poor | 0.2369 | 0.0169 | 0.0019 | 0.0001 | 0.0000 |
| Poor | 0.2981 | 0.0416 | 0.0045 | 0.0007 | 0.0001 |
| Medium | 0.2918 | 0.0746 | 0.0070 | 0.0010 | 0.0003 |
| Good | 0.2323 | 0.0922 | 0.0073 | 0.0011 | 0.0004 |
| Very good | 0.1734 | 0.1017 | 0.0065 | 0.0009 | 0.0002 |

### B. Discussion

The overall result of student performance analysis shows that the domain knowledge influences student performance in general. The pattern of non-negative improvement indicates that more students are likely to perform as expected if they possess a more substantial knowledge on the course topic. Such knowledge also helps them to perform beyond expectation. The growing confidence of positive improvement in the observation supports this fact. However, the positive improvements for very good students are comparatively less than other students. The limited scope of positive improvement for the very good student can be one of the reason behind this.

*1) Measurement of domain knowledge:* Researchers have earlier tried to quantify domain knowledge by using various methods. In one such study [29], the authors have used a set of special questionnaire to determine the student cognition. The specialisation of study or work [16] and pre-test score [18] are also used in this regard. [17] has even used the previous performance in related courses to quantify the same. In order to determine related courses, the authors choose a chemistry course for analysis and consider earlier performance in chemistry-related courses.

The use of a special questionnaire for measuring domain knowledge before course commencement requires student participation, and it is not feasible always. Similarly, the specialisation of study or work may not be helpful as well. It may be possible that a student belongs to other specialisation and attends similar courses earlier. The knowledge is expected to be high in such cases. Obtaining the pre-test score is also not feasible in all cases. Furthermore, the method proposed in [17] does not quantify the domain knowledge in general; instead, it chooses the courses manually.

In contrast, the proposed approach does not require any questionnaire for the students or an arrangement of a pre-test. It does not even require any manual intervention as the course syllabus is widely available in digital format nowadays. Moreover, many institutes have started maintaining course registration and performance records online. The proposed approach is capable of measuring the domain knowledge from the data stored in the information system of an institute. The present work advances the method proposed in [17] by applying a semantic approach to measure the similarity between courses. Such similarity measures, along with the historical performance and time dimension, finally helps in quantifying the domain knowledge automatically without any manual intervention.

*2) Domain knowledge: a potential predictor:* In order to build a prediction model, the set of predictor should be identified first, and it is valid for any predictive task. In a previous study, [10] mentioned that the CGPA and internal assessments are sufficient enough to predict the performance of a student. Many previous studies have, in fact, used the CGPA and internal assessment marks as the predictor and finally achieved significant prediction accuracy [9]. However, the accuracy increases towards the end of the course when more internal assessment marks are available [10]. Unfortunately, these internal assessment marks are available only during the tenure of the course, and they can not be used for early prediction before course commencement. Moreover, the existing early prediction studies are not efficient enough in terms of their accuracy and therefore emerging as a future research direction [5]. The unavailability of effective predictor could be one of the reasons behind this. Thus, it is important to identify some more predictors which can be measured early.

The present work has explored the potential of domain knowledge for this purpose. The in-depth analysis shows that domain knowledge has a positive influence on student performance. This work has proposed an approach which measures the domain knowledge based on the course syllabus, performance in previously attended courses, and the time gap between course registration. Importantly, this approach can facilitate the domain knowledge measurement before the course commencement as the syllabus, past performance and registration schedule are all known apriori. The domain knowledge, measured using the proposed approach, can be utilised as a predictor in performance prediction studies where the objective is to predict performance before the course commencement.

## V. CONCLUSION

The present work has applied advanced data mining technologies on a sufficiently large dataset and tried to explore the association between domain knowledge and student performance. It initially applies a semantic method on the course syllabus to find the similarity between courses. These similarity measures further help in quantifying the domain knowledge based on the performance in similar courses attended recently. The association mining finally establishes the influence of domain knowledge on student performance.

Although previous studies have tried to quantify the domain knowledge based on the questionnaire, pre-test score or by manually selecting courses, these methods are not feasible for application in an automated setting. The present work has primarily addressed this concern. As a primary contribution, it has proposed a semantic approach which facilitates the measurement of domain knowledge based on available data only. Besides this, further analysis helps in understanding an important pedagogical aspects related to student performance. The performance analysis establishes the fact that sufficient domain knowledge not only motivates a student to perform as per expectation but also help to perform beyond that.

As the present study has already established a positive association here, future studies on student performance prediction can utilise the domain knowledge measured using the proposed method. This work has applied the semantic method proposed in [24] for measuring the similarity between courses as it was earlier used in several educational studies. However, it does not compare other available semantic methods. A more promising similarity measurement can yield a stronger association between domain knowledge and student performance. As the proposed approach is capable of adapting other methods of similarity calculation, the authors, therefore, have a plan to carry out an effectiveness comparison in future. Besides this, the proposed approach of domain knowledge measurement can be evaluated with the dataset of other institutes for establishing its' validity further. The comparison study with conventional approaches can be another interesting direction.